\documentclass[12pt]{article}

\usepackage{microtype}
\usepackage{graphicx}
\usepackage[dvipsnames]{xcolor}
\usepackage[body={6.6in, 9in}, right=1in, top=1in]{geometry}
\usepackage{amssymb}
\usepackage{amsmath} 
\usepackage{feynmp} 
\usepackage{subfigure}
\usepackage[numbers,sort&compress]{natbib}

\definecolor{darkblue}{rgb}{0.15,0.35,0.55}
\definecolor{reddish}{rgb}{0.65, 0.2, 0.2}
\usepackage[linktoc=page,colorlinks=true]{hyperref}
\hypersetup{citecolor=darkblue,
linkcolor=reddish,
urlcolor=darkblue}

\def\be{\begin{equation}}
\def\ee{\end{equation}}
\def\ba{\begin{eqnarray}}
\def\ea{\end{eqnarray}}

\def\le{\left}
\def\ri{\right}
\newcommand\ov{\over}

\newcommand{\es}[2] {\begin{equation} \label{#1} \begin{split} #2 \end{split} \end{equation}}
\def\<{\langle}
\def\>{\rangle}

\newcommand\sig{\sigma}

\newcommand\ga{{\ensuremath{{\gamma}}}}

\newcommand\snowmass{\begin{center}\rule[-0.2in]{\hsize}{0.01in}\\\rule{\hsize}{0.01in}\\
\vskip 0.1in Submitted to the  Proceedings of the US Community Study\\ 
on the Future of Particle Physics (Snowmass 2021)\\ 
\rule{\hsize}{0.01in}\\\rule[+0.2in]{\hsize}{0.01in} \end{center}}

\begin{document}

\thispagestyle{empty}


\begin{center}
\Large{\textbf{Snowmass White Paper: \vspace{.15cm} \\ Effective Field Theories for Condensed Matter Systems}} \\[1cm]
\large{Tomas Brauner$^{\rm a}$, Sean A. Hartnoll$^{\rm b}$, Pavel Kovtun$^{\rm c}$, Hong Liu$^{\rm d}$, Mark Mezei$^{\rm e}$, \vspace{.15cm} \\ Alberto Nicolis$^{\rm f}$, Riccardo Penco$^{\rm g}$, Shu-Heng Shao$^{\rm h}$, Dam Thanh Son$^{\rm i}$}
\\[.8cm]

\vspace{.2cm}
\small{\textit{$^{\rm a}$Department of Mathematics and Physics, University of Stavanger, N-4036 Stavanger, Norway}}

\vspace{.2cm}
\small{\textit{$^{\rm b}$DAMTP, University of Cambridge, Cambridge CB3 0WA, UK}}

\vspace{.2cm}
\small{\textit{$^{\rm c}$Department of Physics \& Astronomy, University of Victoria, Victoria, BC, V8W 2Y2, Canada}}

\vspace{.2cm}
\small{\textit{$^{\rm d}$Center for Theoretical Physics,
Massachusetts Institute of Technology, Cambridge, MA 02139, USA}}

\vspace{.2cm}
\small{\textit{$^{\rm e}$Simons Center for Geometry and Physics, SUNY, Stony Brook, NY 11794, USA}}

\vspace{.2cm}
\small{\textit{$^{\rm f}$Center for Theoretical Physics, Columbia University, New York, NY 10027, USA}}

\vspace{.2cm}
\small{\textit{$^{\rm g}$Department of Physics, Carnegie Mellon University, Pittsburgh, PA 15213, USA}}

\vspace{.2cm}
\small{\textit{$^{\rm h}$C. N. Yang Institute for Theoretical Physics, Stony Brook University, Stony Brook, NY 11794, USA}}

\vspace{.2cm}
\small{\textit{$^{\rm i}$Kadanoff Center for Theoretical Physics, University of Chicago, Chicago, IL 60637, USA}}

\end{center}

\vspace{.3cm}


\begin{abstract}
We review recent progress and a number of future directions for applications of effective field theory methods to condensed matter systems broadly defined. Our emphasis is on areas that have allowed a fertile exchange of ideas between high energy physics and many-body theory. We discuss developments in the effective field theory of spontaneous symmetry breaking, of hydrodynamics and non-equilibrium dynamics more generally, fracton phases of matter, and dualities between 2+1 dimensional field theories. We furthermore discuss the application of effective field theory to non-Fermi liquids, the dynamics of entanglement entropy and to condensed matter aspects of cosmology.
\end{abstract}

\snowmass

\newpage

\setcounter{page}{1}

\tableofcontents

\section{Introduction}

High energy particle physics is described by a Lorentz invariant quantum field theory \cite{Weinberg:1995mt}. Taking this high energy point of view literally, condensed matter physics is concerned with certain nontrivial states of this quantum field theory that contain a nonzero density of particles. These could be solids, fluids, clouds of cold atoms, etc. Crucially, these states necessarily break Lorentz invariance while often preserving some notion of translational invariance. This symmetry-based perspective allows a powerfully general classification of forms of condensed matter \cite{Nicolis:2015sra}, and remnants of the full symmetry can be present in the form of Golstone bosons.

At energies that are low compared to particle physics scales, a given state of condensed matter can be described starting from an effective low-energy Hamiltonian with no vestige of the high energy symmetries. This description may take the form of a lattice model describing, for example, the interactions of mobile electrons with lattice vibrations. The only spacetime symmetries of such a model will be discrete translations and rotations. That is to say, this description is not necessarily field theoretic. However, a further low energy limit can be taken, which considers physics on energies well below the lattice energy scale. It is found that in many important circumstances, universal continuum physics re-emerges. This occurs in gapless systems such as quantum critical points and Fermi liquids, in gapped systems with long-range topological interactions, and in nonzero temperature systems described by some form of hydrodynamics. This emergent universality is captured by effective field theories. 
There are also notable exceptions to this paradigm, where the low-energy physics is not captured by a conventional continuum field theory, such as in the case of fractons.

The particles that arise in condensed matter effective field theories are distinct from the microscopic particles of the Standard Model. For this reason they were called `quasiparticles' by Landau.
The quasiparticles can be almost free or strongly interacting. Furthermore, these field theories are generically not Lorentz invariant (although they can be). The absence of Lorentz invariance means that condensed matter field theories are far less constrained than their high energy cousins and frequently push the boundaries of our understanding of what quantum field theory can do. For example, the structure of ultraviolet and infrared divergences can be much richer. Nonetheless, historically, key notions in field theory including the development of the renormalization group, the importance of spontaneous symmetry breaking and the role of topological excitations were developed with significant synergy between high energy and condensed matter physics.

The last two decades have seen much progress in the development of effective field theories for condensed matter systems, with an ongoing interchange of ideas between high energy and condensed matter physics. Here we highlight a number of areas which look promising for substantial progress in the next decade or so. Our list is meant to be illustrative, and by no means exhaustive.

\section{Spontaneous Symmetry Breaking} 
\label{sec:SSB}

Spontaneous symmetry breaking (SSB) is a classic subject covered in most textbooks on quantum or many-body field theory. Early progress in the field occurred in parallel in particle and condensed matter physics. However, the subsequent development of effective field theory (EFT) techniques for SSB was motivated largely by spontaneous breaking of internal symmetry in relativistic field theory~\cite{Coleman1969a,Callan1969a,Volkov1973a}. This remained the state of the art until the turn of the century. Around two decades ago, the growing number of novel applications to dense relativistic matter as well as to condensed matter physics led to a revival of interest in the general properties of SSB. This culminated in a general classification of Nambu--Goldstone (NG) bosons~\cite{Watanabe2011a,Watanabe2012b,Hidaka2013b, Nicolis:2013sga}, and a general low-energy EFT description of NG bosons~\cite{Leutwyler1994b,Watanabe2014a}, in quantum many-body systems.

As the above suggests, application of EFT to the low-energy description of quantum matter is an area where the high energy physics community can contribute expertise accumulated over the last half century. At the same time, the explosive rate of development in contemporary condensed matter and material physics provides a rich source of novel applications, stimulating further development of the theory. However, in order for the knowledge transfer to be efficient, it is necessary to make the EFT formalism for SSB as ``user-friendly'' as possible. This goal has so far been achieved only for SSB of internal symmetries. When it comes to SSB of spacetime symmetry, in particular translations, much progress has already been made (see e.g.~\cite{Watanabe2013a,Nicolis2014a,Hidaka2015a,Horn:2015zna,Nicolis:2017eqo,Pavaskar:2021pfo} for some examples). However, the construction of effective actions still requires a substantial amount of insight. Setting up an algorithmic guide that automatizes the path from the symmetry-breaking pattern to the effective action should be an important goal for the future application-oriented research.

On the more conceptual side, a question that has been gaining importance is how to identify and choose the infrared data necessary for the construction of the low energy EFT for a given system. The traditional approach of relativistic EFT has been to identify the symmetries present in the ultraviolet as well as the pattern of their spontaneous breaking in the infrared~\cite{Coleman1969a,Callan1969a,Volkov1973a}. However, already the early work on EFT for systems without Lorentz invariance~\cite{Leutwyler1994b,Watanabe2014a} showed that additional infrared data, such as the vacuum expectation values of conserved charges, may be needed. More recently, research on scattering amplitudes has led to the proposal to construct low energy EFT using \emph{solely} infrared data, without reference to the spontaneously broken symmetry, manifest in the ultraviolet. Thus, the EFT for NG bosons can be recovered if one knows the spectrum of massless particles, the linearly realized (unbroken) symmetry, and the soft behavior of scattering amplitudes of NG bosons~\cite{Low2015a,Cheung2015a}.

In addition, low-temperature phases of matter can come with new, emergent symmetries. These may be (typically approximate) ``accidental'' symmetries associated with a noninteracting infrared fixed point. Alternatively, one may find generalized, higher-form symmetries, typically associated with the presence of topological excitations in the system~\cite{Gaiotto:2014kfa}. The presence of higher-form symmetries of different degrees may in turn lead to novel algebraic structures beyond ordinary group theory~\cite{Cordova2019a,Hidaka2020d,Brauner2021b}. In the last couple of years, yet another class of generalized symmetries and associated conservation laws has proven essential for understanding the physics of exotic quantum phenomena such as fractons and the fractional quantum Hall effect~\cite{Gromov2019a,Seiberg2020a,Du2021a}. Since conservation laws largely dictate the low-energy physics of systems with SSB, it is imperative to gain deeper, systematic understanding of the origin of the various emergent symmetries and their interplay.


\section{Non-equilibrium systems}
\label{sec:noneq}

Recently there has been significant progress in developing EFTs for describing large-distance and long-time behavior of a strongly coupled quantum many-body system in far-from-equilibrium states (for a review see~\cite{Liu:2018kfw}).
 There are a number of new elements in the formulation of  non-equilibrium EFTs
  compared with that of the more familiar ones for systems in equilibrium or vacuum states. Firstly, the dynamical variables are associated with conserved quantities and Goldstone modes,
    as non-conserved quantities in general relax to local equilibrium in microscopic time scales.\footnote{With the exception that order parameters should also be included in the non-equilibrium dynamics if the system is near a continuous phase transition.} 
Secondly, such EFTs are formulated on a Schwinger-Keldysh contour (or closed time path) which means that the dynamical variables are doubled. In particular, unitarity of quantum evolution on a closed time path imposes various constraint conditions  that persist in the classical limit. Thirdly, they incorporate dissipative terms that violate time reversal macroscopically, macroscopic implications of microscopic time reversal symmetry, and local fluctuation-dissipation relations in a unified manner. 

An immediate application of the non-equilibrium EFT formalism is a new formulation of hydrodynamics based solely on symmetries and action principle~\cite{Crossley:2015evo,Glorioso:2017fpd} (see also~\cite{Kovtun:2014hpa,Harder:2015nxa,Haehl:2015pja,Haehl:2015uoc,Jensen:2017kzi} and some earlier work includes~\cite{Dubovsky:2005xd,Dubovsky:2011sj,Nicolis:2013lma,Endlich:2010hf,Endlich:2012vt,Grozdanov:2013dba,Delacretaz:2014jka}). 
This formulation led to a new derivation of the second law of thermodynamics, including a universal formula for the amount of entropy production~\cite{Glorioso:2016gsa}. The hydrodynamic action makes it possible to calculate systematically 
universal corrections from hydrodynamic fluctuations to correlation functions of the stress tensor of a quantum many-body system near equilibrium (see e.g.~\cite{Chen-Lin:2018kfl,Delacretaz:2020nit, Lau:2019slu,Jain:2020zhu,Jain:2020hcu}). 
The new formulation also offers a transparent way to derive hydrodynamic descriptions of exotic condensed matter systems where the conventional phenomenological approach does not work. Examples include fracton hydrodynamics for systems with dipole conservation~\cite{Gromov:2020yoc}, hydrodynamics for systems with discrete rotational symmetries~\cite{Huang:2022ixj}, and a new formulation of magnetohydrodynamics for neutron stars~\cite{magneto}. It has also been used to derive a chaos effective theory for maximally chaotic systems~\cite{Blake:2017ris}, and the behavior of spectral form factors for general quantum chaotic systems~\cite{winer2021hydrodynamic}.

The non-equilibrium EFT formalism potentially opens up new avenues for treating a large number of non-equilibrium systems in diverse disciplines.  It has been used to derive effective actions for other continuous media including solids, liquid crystals, and fluids with chemical reactions~\cite{Landry:2019iel,Landry:2020tbh,Baggioli:2020haa,Landry:2021kko}, and for topological Floquet systems~\cite{Glorioso:2019koh}. 
It can be used to study systematically the interplay between hydrodynamic and order parameter fluctuations on dynamical critical phenomena for systems near a critical point, a question of great interests in the ongoing experimental program searching for the QCD critical point~\cite{Luo:2017faz}, as well as many condensed matter contexts~\cite{hohenberg1977theory}. We expect it to also have applications to effective descriptions of formation of large scale structure of the universe~\cite{Baumann:2010tm},  black hole mergers~\cite{Goldberger:2004jt}, driven open systems~\cite{Sieberer:2015svu}, as well as characterizing entropy production fluctuations~\cite{crooks1999entropy}.


\section{Hydrodynamics}

Among all possible out-of-equilibrium phenomena, hydrodynamic behavior deserves a special mention given how frequently it is encountered in Nature. Hydrodynamics has been historically formulated as a classical theory, whose subtleties largely lie in the non-linear nature of the differential equations, subject to the appropriate initial and boundary conditions. Over the last decade or so, many fundamental questions in hydrodynamics have been revisited thanks to an influx of ideas initially developed for effective field theory (EFT) in high-energy physics. Traditional EFTs have been developed in order to describe fluctuations on top of a simple state, such as the vacuum. On the other hand, hydrodynamics is {\em the} EFT for macroscopic fluctuations on top of a (locally) equilibrium macro-state. As a result of applying EFT methods to hydrodynamics, important insights have been gained. These insights relate both to the classical hydrodynamic theory, and to the broader theoretical picture in which classical hydrodynamics emerges as a ``saddle point'' of a more fundamental description. 

Most of the comments below concern simple hydrodynamics of relativistic theories due to its closer connection with high-energy physics. Many insights have been ported to theories with extra Goldstone modes (such as superfluids), and to Galilean theories or theories with no boost symmetry at all. Indeed, there is a close connection between hydrodynamics and the discussion of spontaneous symmetry breaking in section \ref{sec:SSB}. For example, \cite{Delacretaz:2019wzh} developed a hydrodynamic theory of magnetophonons based on the theory of noncommuting Goldstone bosons \cite{Watanabe2012b}. The ability to explicitly realize hydrodynamic regimes in holographic models has furthermore led to a major effort to systematically incorporate spontaneous breaking of spacetime symmetries in hydrodynamics, recently reviewed in \cite{Baggioli:2022pyb}.

Some of the EFT-inspired insights in classical hydrodynamics include: Derivative expansion in hydrodynamics~\cite{Baier:2007ix, Bhattacharyya:2007vjd} including convergence of the expansion in solvable models~\cite{Heller:2013fn, Grozdanov:2019kge}; Chiral anomalies in hydrodynamics~\cite{Son:2009tf} and the associated physical effects such as the chiral magnetic effect~\cite{Zakharov:2012vv}; New constraints on transport parameters from the existence of generating functionals~\cite{Banerjee:2012iz, Jensen:2012jh}; Classification of transport coefficients in hydrodynamics~\cite{Haehl:2015pja}; The use of generalized symmetries~\cite{Gaiotto:2014kfa} to formulate magneto-hydrodynamics~\cite{Grozdanov:2016tdf, Hernandez:2017mch}; Well-posedness and stability of hydrodynamic equations through higher-derivative regularizations~\cite{Bemfica:2017wps, Kovtun:2019hdm}. 

One marked difference between hydrodynamics and a more traditional EFT has been the apparent lack of an action principle in order to describe the physics of dissipation. While simple actions for non-dissipative fluids~\cite{Dubovsky:2011sj, Bhattacharya:2012zx} and for dissipative fluids~\cite{Kovtun:2014hpa} are based on seemingly different approaches, the two can be unified in the Schwinger-Keldysh construction discussed in Section \ref{sec:noneq}. With a true EFT for thermal states based on the path integral over the emergent low-energy degrees of freedom, the way is open both for formal developments such as~\cite{Jensen:2017kzi, Haehl:2018lcu} and
the systematic incorporation of fluctuation effects in physical correlation functions~\cite{Kovtun:2012rj, Chen-Lin:2018kfl}. The possibilities allowed by these developments are illustrated in the 1+1 dimensional edge theories of quantum Hall states. There, the interplay of anomalies and hydrodynamic fluctuations fundamentally alters the nature of transport ~\cite{Delacretaz:2020jis}. Given the EFT action, one can identify coupling constants which are invisible in classical hydrodynamics, and at the same time contribute to observable low-energy correlation functions through loops~\cite{Jain:2020zhu}. With the above developments, the word ``hydrodynamics'' has essentially become a shorthand for Schwinger-Keldysh EFT describing near-thermal states. The EFT approach to hydrodynamics is in its infancy, with experimentally relevant problems such as critical dynamics in strongly non-equilibrium states still awaiting an adequate EFT treatment.


\section{Large $N$ approaches to Non-Fermi liquids}

Large $N$ limits of fields theories allow the controlled resummation of certain classes of Feynman diagrams, with other processes suppressed by inverse powers of $N$ \cite{Coleman:1980nk}. This gives insight into strong coupling regimes of field theory that are beyond simple perturbation theory. An important example is the Wilson-Fisher fixed point of the $O(N)$ model -- this is a Lorentz-invariant theory that describes classical and quantum phase transitions in magnets and superfluids \cite{ssbook}.

A nonzero density of fermions leads to a Fermi surface, with low energy excitations supported on a nontrivial locus in momentum space (away from $\vec k=0$). Effective field theories describing the low energy dynamics of Fermi surfaces were developed by Shankar \cite{shankarrg} and Polchinski \cite{Polchinski:1992ed}. However, when the fermions have Yukawa-like couplings to additional massless bosonic modes these theories are found to run to strong coupling. The fate of strongly coupled Fermi surfaces is a major open problem in quantum field theory, and plausibly underpins the unconventional `non-Fermi liquid' behavior observed in condensed matter systems where a Fermi surface coexists with a critical bosonic mode \cite{Sachdev:2011cs}. These regimes may also be relevant to high density quark matter \cite{PhysRevD.59.094019}. The early Hertz-Millis-Moriya theory essentially coupled $O(N)$-type models to a single Fermi surface, integrated out the fermions and solved the resulting bosonic theory at large $N$ using established methods. This is not controlled because non-analyticities arise upon integrating out the gapless fermions. More recent work overcame this difficulty by keeping the fermions in the critical theory \cite{PhysRevB.82.075127, PhysRevB.82.075128}. However, it was subsequently shown that these theories contain IR divergences at high loop order that signal the breakdown of the large $N$ expansion \cite{PhysRevB.80.165102}. This observation has triggered a large and ongoing effort to find new large $N$ limits for these systems, that resum different sets of diagrams. We restrict ourselves to aspects of this endeavor that have brought together ideas from both high energy and condensed matter theory.

One approach is to use matrix rather than vector large $N$ theories. Such theories are familiar from large $N$ approaches to QCD. In their application to the non-Fermi liquid problem the fermions are taken in the fundamental representation of $SU(N)$ and the bosonic field in the adjoint \cite{PhysRevLett.123.096402}. Intriguingly, superconductivity emerges from this field-theoretic model \cite{PhysRevB.92.205104} in a way that is very similar to holographic models of non-Fermi liquids \cite{hartnoll2018holographic}. The holographic models also describe matrix large $N$ theories and map non-Fermi liquid scaling laws onto the near-horizon behavior of charged black holes. Again closely related to charged black holes is the SYK \cite{SY92, kitaev2015talk} class of large $N$ theories of non-Fermi liquids. The essential technical advance here is that allowing certain couplings to be drawn from a random ensemble simplifies the large $N$ Feynman diagrammatics while preserving the core physics. This rapidly evolving field has recently been reviewed in 
\cite{chowdhury2021sachdevyekitaev}.

A fundamentally different type of large $N$ approach to non-Fermi liquids involves discretizing the Fermi surface itself into `patches' that effectively decouple at low energies. This leads to a very large low energy symmetry group that describes membrane-like dynamics for the Fermi surface \cite{PhysRevLett.72.1393}. This construction has been refined and generalized very recently, with the emergent dynamics constrained through 't Hooft anomaly matching \cite{PhysRevX.11.021005}
and a fully nonlinear effective field theory of the Fermi surface membrane developed
\cite{Delacretaz:2022ocm}. Future work will determine whether this provides a controlled framework for non-Fermi liquids, perhaps even allowing a systematic classification of strongly coupled Fermi surfaces.


\section{Dualities in $2+1$ dimensions} 

Duality has been an important tool in theoretical physics. A duality establishes an equivalence between distinct effective field theories, thereby connecting disparate microscopic systems and sometimes allowing computations in strongly coupled regimes by using a simpler dual description. Starting
from the Kramers-Wannier duality in the Ising model, the late
1970's--early 1980's saw the discovery of the duality between the abelian
Higgs model and the $O(2)$ scalar field theory (the ``bosonic
particle-vortex duality''), with important consequences for the
superconductor-insulator phase transition in condensed matter
physics~\cite{Peskin:1977kp,Dasgupta:1981zz}.  More recently, examples
of duality between large-$N$ bosonic and fermionic Chern-Simons field
theories have been found~\cite{Aharony:2015mjs}.  This indicates that
duality is a rather widespread phenomenon in nonsupersymmetric
(2+1)-dimensional field theory.

The newest impetus to the discovery of new duality pairs comes from the
physics of the fractional quantum Hall effect, which occurs when a
two-dimensional electron gas is subjected to a magnetic field large
enough so that all of the electrons are on the lowest Landau level.
When the Landau level is half full, the resulting state is a Fermi
liquid of the so-called ``composite fermions.''  For a long time, the
standard theory of the composite fermion was the Halperin--Lee--Read
theory, in which the composite fermion is interpreted as an electron
with two flux quanta attached.  Despite some phenomenological success,
the theory has a serious problem---it breaks a discrete symmetry of
the half-filled Landau level---the particle-hole symmetry.

The recently proposed solution to the problem of particle-hole
symmetry, the Dirac composite fermion theory, postulates a duality
between the theory of a free fermion and that of a fermion coupled to a gauge
field (the ``fermionic particle-hole duality'')~\cite{Son:2015xqa}.
This duality turns out to have important consequences for the
interacting surface of a three-dimensional topological
insulator~\cite{Wang:2015qmt,Metlitski:2015eka}.  Later it was found
that both the bosonic and fermionic particle-hole dualities can be
derived from a more elementary ``seed duality,'' from which a whole
web of new dualities can be derived~\cite{Karch:2016sxi,Seiberg:2016gmd}.

It is fair to say that our knowledge of dualities among
(2+1)-dimensional quantum field theories is still rudimentary, and their
study has only just begun.


\section{Exotic Field Theories for Fractons}

Starting from a microscopic lattice model with local interactions,  it is generally believed that there is a low-energy continuum field theory description at long distances. 
For instance, in three spacetime dimensions, most, if not all, gapped lattice models admit a description in terms of a Topological Quantum Field Theory, such as the Chern-Simons gauge theory, in the low energy limit. 
This piece of lore  has been challenged by a novel class of lattice models (mostly in four spacetime dimensions) known as \textit{fractons} \cite{PhysRevLett.94.040402,PhysRevA.83.042330}. 
See \cite{Nandkishore:2018sel,Pretko:2020cko} for reviews. 
The key characteristics of the fracton models include:
\begin{itemize}
\item The ground state degeneracy of these models depends sensitively on the number of  lattice sites, and becomes infinite in the continuum limit. Furthermore, this large  degeneracy is robust and cannot be lifted by local operators in perturbation theory.  
\item The spectrum  includes particle excitations with restricted mobility. Some of them cannot move in isolation, while some others can only move along certain directions in space.
\end{itemize}
These peculiarities seem impossible to fit into a conventional field theory framework.

There has been a lot of exciting developments in understanding these peculiar properties in the continuum limit, which typically involves certain exotic field theories. 
In \cite{Pretko:2016kxt,Pretko:2016lgv,Pretko:2018jbi}, it was pointed out that the restricted mobility of the fracton particles can be realized from a \textit{higher-rank  gauge theory} (which was originally studied in a different context in \cite{Gu:2006vw,Xu:2006,Pankov:2007,Xu2008,Gu:2009jh,rasmussen}). 
In higher-rank gauge theory, the spatial components of the gauge fields are not vectors, but higher-rank tensors of the spatial rotation group. 
The gauge transformation of these higher-rank gauge theories typically involve more than one spatial derivative. 
The  fracton particles are charged under the higher-rank gauge symmetry, and their motion is constrained by the multipole moment conservation from the gauge symmetry. 
More precisely, the restricted mobility can be explained as a consequence of a global symmetry that acts on the Wilson defects, which represent the worldlines of infinitely heavy fracton particles, in higher-rank gauge theory  \cite{Gorantla:2022eem}. 
See \cite{Slagle:2017wrc,Prem:2017kxc,2018PhRvB..98c5111M,Bulmash:2018lid,2018PhRvB..98l5105M,Slagle:2018kqf,2018PhRvL.121w5301P,Williamson:2018ofu,2020PhRvR...2b3249Y,You:2019bvu,Radzihovsky:2019jdo,Seiberg:2019vrp,Wang:2019aiq,Wang:2019cbj,Gromov:2020rtl,Gromov:2020yoc,paper1,paper2,paper3,Fontana:2020tby,Slagle:2020ugk,Karch:2020yuy,Qi:2020jrf,Yamaguchi:2021qrx,Fontana:2021zwt,Du:2021pbc,Hsin:2021mjn,Jain:2021ibh} for other developments in higher-rank gauge theory.

The key common feature of many of the fracton models  is the existence of the \textit{subsystem global symmetry},  generalization of  ordinary global symmetries. 
It is a symmetry that acts only on a subspace of the whole system. 
Unlike the higher-form global symmetry \cite{Gaiotto:2014kfa}, the   subsystem symmetry charge depends not only on the topology, but also on the detailed shape of the subspace and its location. 
In the continuum limit, the number of independent conserved charges typically diverges. 
This fact makes the long-distance behavior of these systems sensitive to certain short-distance details.
This  surprising   UV/IR mixing \cite{paper1,Gorantla:2021bda,You:2021tmm,Zhou:2021wsv,You:2021sou} is the underlying reason making these models defy a conventional field theory description.

A systematic investigation of these new field theories with subsystem symmetries was done in a series of papers \cite{paper1,paper2,paper3,Gorantla:2020xap,Gorantla:2020jpy,Karch:2020yuy,Rudelius:2020kta,Gorantla:2021svj,Gorantla:2021bda,Burnell:2021reh,Distler:2021qzc,Distler:2021bop}.  
They studied many higher-rank tensor gauge theories of fractons as well as other models such as the one from \cite{PhysRevB.66.054526}.  
Important global aspects of these exotic field theories, such as the compactness of the fields,  the geometric structure of the bundles for the tensor gauge fields, and the anomalies of subsystem symmetries, were thoroughly analyzed in a generalized framework of continuum field theory.

One of the simplest gapped fracton models is the X-cube model, first introduced in \cite{Vijay:2016phm}. 
Its ground state degeneracy is $2^{2L_x+2L_y+2L_z-3}$, where $L_i$ is the number of lattice sites in the $i$ direction.  
A field theory description for the X-cube model was developed in \cite{Slagle:2018kqf,paper3}, which takes the form  as a BF-type Lagrangian of tensor gauge fields.  
However, unlike the usual topological BF theory and the familiar Chern-Simons gauge theory, the X-cube field theory is not topological; in fact, it is not even invariant under continuous spatial rotation. 
The X-cube field theory has discrete subsystem global symmetries with an 't Hooft anomaly, which leads to the large ground state degeneracy \cite{paper3,Burnell:2021reh}. 
See also \cite{Slagle:2018swq,Aasen:2020zru,Slagle:2020ugk,Hsin:2021mjn,Geng:2021cmq} for alternative field theory descriptions of the X-cube model and related fracton models. 

Despite progress in the development of nonstandard field theories for some of the simpler fracton models, there is still no  continuum field theory framework for the more general models. In particular, it remains an outstanding open question to construct a continuum field theory description for one of the first fracton models, the Haah code \cite{PhysRevA.83.042330}. (See, for example,  \cite{Bulmash:2018knk,Gromov:2020rtl,Fontana:2021zwt} for developments in this direction.) 
Building  a new framework of continuum  field theory to incorporate these fracton phases of matter is important  for understanding the universal properties of these models and for finding an organizing principle for classifying them. 
More intriguingly, it will also guide us towards possible generalizations of standard quantum field theory.


\section{Membrane dynamics of entanglement entropy}

Quantum entanglement is a central and unifying concept in many branches of quantum mechanics: from quantum information theory, through condensed matter theory, to quantum gravity. Some of these ideas have been discussed in separate white papers \cite{Bousso:2022ntt, Blake:2022uyo}. 
Here we focus on universal aspects of quantum entanglement dynamics that underpin the emergence of many-body statistical mechanics and which may be captured by a membrane effective field theory.  

Let us consider a pure initial state that has negligible entanglement between subregions. 
The state evolves unitarily, developing significant entanglement between its subsystems.
While the full system remains in a pure state with zero entropy, at late times local subsystems are described by thermal mixed states with their entanglement entropy (EE) playing the role of the thermal entropy of statistical physics. Hence, the time evolution of EE is a sensitive diagnostic of the emergence of statistical mechanics. 

A tremendous amount of effort has gone into experimentally measuring  EE~\cite{2016exp,2019exp,2019exp2} and developing computational tools for it: notable examples are tensor networks and entanglement renormalization~\cite{PhysRevLett.69.2863,2005,Vidal:2007hda}, the replica trick~\cite{Holzhey:1994we,Calabrese:2004eu,Calabrese:2005in} and the holographic entropy formula~\cite{Ryu:2006bv,Ryu:2006ef,Hubeny:2007xt,Lewkowycz:2013nqa}. An interesting question is whether there is any universality in the dynamics of EE across different quantum systems. A promising kinematic regime is that of large subregions and late times, where the leading behavior of EE is expected to behave as:
\es{Scaling}{
S_\text{EE}[A(t)]=s_\text{th}R^{d-1}\, {\cal S}\le(t/ R\ri)\le[1+{\cal O}(\beta/R)\ri]\,, \qquad \text{for}\quad t,R\gg \beta\,, 
}
where we take the subregion $A$ with characteristic size $R$ to lie on the fixed $t$ time slice, $s_\text{th}$ is the thermal entropy density of the system at the effective inverse temperature $\beta$ (that can be read off from the energy density), $d$ is the number of spacetime dimensions, and ${\cal S}$ is a scaling function that only depends on the {\it ratio} of the large kinematic data $t$ and $R$. The scaling behavior in eq.~\eqref{Scaling} was indeed verified in a variety of systems~\cite{Calabrese:2005in,2013,Liu:2013iza,Liu:2013qca}. The next step is to contemplate whether there exists an effective field theory (EFT) framework that can predict the scaling function ${\cal S}$.

\begin{figure}[!h]
   \centering
      \includegraphics[width=0.4\textwidth]{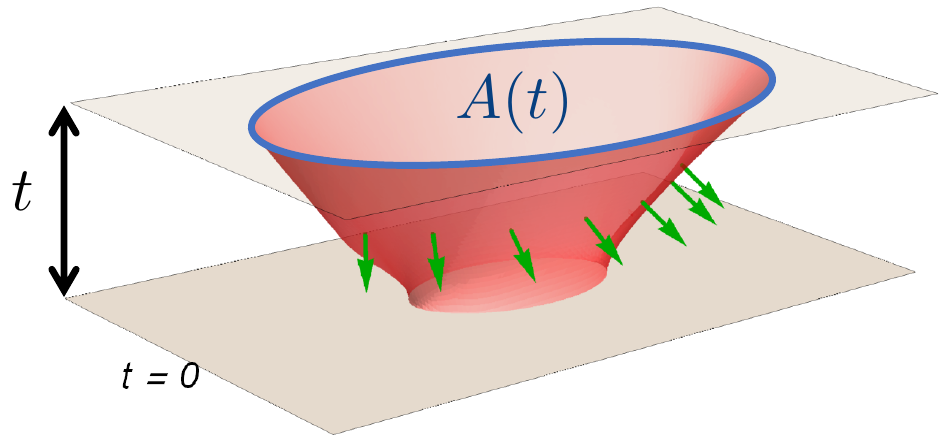}
  \caption{Membrane minimizing the action~\eqref{MembraneAction} for $A(t)$ an ellipse, adapted from~\cite{Mezei:2020knv}. The two faces of the Minkowski spacetime slab represent the initial state at $t=0$, and the time slice at $t$. The green vectors are the local unit normals $n(\sig)$.
  \label{fig:membrane} }
\end{figure}

Based on the study of random quantum circuits~\cite{Jonay:2018yei} and holographic QFTs~\cite{Mezei:2018jco} (building on earlier partial results in~\cite{Nahum:2016muy,Mezei:2016zxg}) an EFT for EE dynamics in the large subregion-late time regime was proposed. Convincing independent numerical evidence for its validity was found in chaotic spin chains~\cite{Jonay:2018yei} and Floquet systems~\cite{Zhou:2019pob}. 
The theory describes a  codimension-$1$ membrane stretching in spacetime between the  boundary of the subregion $A(t)$ and the initial time slice, see Fig.~\ref{fig:membrane}.\footnote{The membrane is a coarse grained minimal cut through the  circuit in the random circuit context, while in holography it is a lightlike projection of the extremal surface computing EE in the gravity description into the boundary spacetime.} EE is computed by the membrane that minimizes the action:
\es{MembraneAction}{
S_\text{EE}[A(t)]&=\min \int d^{d-1}\sig\, \sqrt{-\ga(\sig)}\, s_\text{th} \, {\cal L}(v(\sig))\le[1+{\cal O}(\beta/R)\ri]\,,\\
v(\sig)&\equiv {(n(\sig)\cdot u)\ov \sqrt{1+ (n(\sig)\cdot u)^2}}\,,
} 
where  $\sig$ are the membrane worldvolume coordinates, $\ga(\sig)$ is the induced metric, ${\cal L}(v)$ is the Lagrangian of the theory that only depends on the quantity $v$ (and not its derivatives) that is determined by the angle between the unit normal $n(\sig)$ to the membrane at point $\sig$ and the timelike unit vector $u$; in $2d$ the membrane becomes a worldline and $v(\sig)$ is its velocity. 
The membrane is anchored on the  boundary of the subregion $A(t)$ and is required to have $v=0$ at the initial $t=0$ time surface.
${\cal L}(v)$ is a {\it Wilson function} of the EFT, i.e.~it depends on the microscopic theory and the value of conserved densities, but not on the details of the highly excited state whose EE dynamics we are describing.\footnote{While a function worth of freedom may sound a lot, the situation is no different from hydrodynamics, where the equation of state is a microscopic input. Once ${\cal L}(v)$ is determined (e.g. from the EE evolution for a spherical subregion), it gives infinitely many predictions for the EE of other shapes.} It can be read off from the dual curved geometry in holography, extracted from the random circuit architecture, and in Floquet systems a precision RG-like numerical method was developed to extract it from the microscopic model~\cite{Zhou:2019pob}. 

While we do not presently have an argument leading to the Membrane EFT in generic systems, the theory has geometric
degrees of freedom that can describe the EE dynamics in any local quantum system. It is an encouraging sign that two disparate class of solvable chaotic models, random circuits and holographic gauge theories lead to the same EFT, and through  these two example systems the Membrane EFT fosters connections between condensed matter and high energy approaches to the dynamics of EE. 
The Membrane EFT is a highly flexible framework that has been generalized and applied in a plethora of directions: for non-homogeneous states, the evolution of conserved densities is described by hydrodynamics. 
The Membrane EFT couples geometrically to hydrodynamics: to leading order in the hydrodynamic expansion we simply promote $s_\text{th}\to s_\text{th}(x(\sig)),\, u_\mu\to u_\mu(x(\sig))$ with $x(\sig)$ the spacetime (target space) point at the membrane coordinate $\sig$ and $s_\text{th}(x)$ and $u_\mu(x)$ are the hydrodynamic entropy density and $d$-velocity~\cite{Mezei:2019zyt}.
If instead the initial state itself contains extensive entanglement, the $v=0$ boundary condition at $t=0$ is changed to $v=f(s_\text{init}/s_\text{th})$, where $f$ is determined by ${\cal L}(v)$~\cite{unpublished,Eccles:2021zum}. The Membrane EFT also captures operator entanglement and EE evolution in splitting and joining quenches, by introducing new features into the spacetime that the membrane probes~\cite{Mezei:2019zyt}. Disorder can be incorporated along the lines of~\cite{Nahum:2017xiy}.

Quantum chaotic dynamics can be explored at different time scales using different phenomena and their corresponding probes, hence there ought to be relations between the data describing them. Above, we explained how the simplest manifestation of chaos, hydrodynamic transport of conserved quantities couples to the entanglement membrane. Excitingly, the butterfly velocity $v_B$ that characterizes scrambling of simple local operators into complex nonlocal ones \cite{Hayden:2007cs,Sekino:2008he,Shenker:2013pqa,Roberts:2014isa,Maldacena:2015waa} manifests itself in ${\cal L}(v)$~\cite{Jonay:2018yei,Mezei:2018jco}.\footnote{Namely the values of  ${\cal L}(v_B)$ and ${\cal L}'(v_B)$ are both fixed.} This surprise connection hints at a fascinating interplay between scrambling and EE dynamics that perhaps can be uncovered in the framework of EFTs for both phenomena, see~\cite{Mezei:2016wfz,Zhou:2019pob,Couch:2019zni,unpublished,Eccles:2021zum} for further hints. It is an outstanding challenge to connect EE dynamics and its Membrane EFT description to other signatures and probes of quantum chaos, such as the random matrix statistics of energy levels~\cite{wigner1951statistical,dyson1962statistical,mehta1963statistical,bohigas1984characterization} and matrix elements~\cite{deutsch1991quantum,srednicki1994chaos} and the growth of complexity~\cite{Susskind:2014rva}.

The rich applicability of the Membrane EFT 
proves its robustness and fuels the hope that soon a general physical derivation for it can be found. 
In Floquet systems  the entanglement membrane was shown to separate different pairing patterns of field configurations between forward and backward time evolution~\cite{Zhou:2019pob}. This mechanism may possibly  be generalized  to Hamiltonian systems and QFTs, it has been applied in related contexts~\cite{Liu:2020jsv,2021PhRvX..11b1051G}, and seems to make contact with the approach of refs.~\cite{Stanford:2021bhl,Gu:2021xaj} to scrambling through an instability in the pairing of the segments of multifold closed time path.
Perhaps generalizing the Membrane EFT  to other entanglement measures, such as reflected \cite{Dutta:2019gen,Kudler-Flam:2020url,Akers:2021pvd} and R\'enyi entropies~\cite{Nahum:2016muy,Zhou:2018myl,Dong:2016fnf}, could provide additional hints as to how the membrane emerges from microscopic degrees of freedom. A reformulation in terms of bit threads may also prove to be useful~\cite{Freedman:2016zud,Agon:2019qgh}. It remains a challenge to understand how to incorporate $1/N$ corrections in holography~\cite{Faulkner:2013ana} and diffusive time dependence of EE in systems with small on-site Hilbert space and a conserved $U(1)$ charge~\cite{2019,2019B,Huang:2019hts} into the Membrane EFT framework.


\section{Condensed matter and cosmology}

An improved understanding of condensed matter systems based on EFTs can also further our understanding of the universe at the largest observable scales. This is because, on the one hand, cosmology stems from the interplay between the condensed matter content of the universe and gravity; on the other hand, experiments aimed at testing the dark matter and dark energy sectors often rely on condensed matter systems with non-trivial properties (e.g. superfluidity, superconductivity, Bose-Einstein condensation, ...).

The large scale behavior of our universe is accurately described by a Friedmann-Robertson-Walker solution to Einstein's equations. This solution, and small perturbations around it, are sourced by an approximately isotropic and homogeneous stress-energy tensor $T_{\mu\nu}$. During most of the history of the universe, this $T_{\mu\nu}$ describes an ordinary viscous fluid with a small pressure~\cite{Baumann:2010tm}. At smallest scales or at early times, however, compelling models have been proposed where the dominant contribution to $T_{\mu\nu}$ comes from matter that is not in a fluid phase. This can lead to qualitatively different predictions for our observable universe. 

For example, an early phase of accelerated expansion---inflation---could have been driven by a solid under some (technically) natural assumptions about its equation of state~\cite{Gruzinov:2004ty,Endlich:2012pz,Kang:2015uha,Nicolis:2020rqz}. This scenario contradicts widely accepted expectations based on traditional models of inflation---e.g. scalar and tensor perturbations are not exactly conserved outside the horizon, and non-gaussian correlation functions of perturbations don't have to be isotropic---while remaining compatible with current observations. More exotic phases of matter in the early universe have also been considered~\cite{Piazza:2017bsd}, but it is fair to say that a systematic exploration of different phases is still lacking.

At smaller scales, where non-linearities become important and gravitational collapse leads to significant matter overdensities,  the dark matter sector could be in a qualitatively different state.
One compelling possibility that has been extensively explored is that dark matter could be in a superfluid state at the center of galaxies~\cite{Berezhiani:2015bqa}. This could explain the observation of galactic scaling relations~\cite{McGaugh:2000sr,mcgaugh2016radial} without the need for ad-hoc modifications of gravity. Once again, a study of the  implications of different phases for the dark matter sector is still in its infancy, and will require EFT techniques to be carried out systematically. 

The superfluidity scenario mentioned above requires a dark matter mass $m \lesssim$ eV. Dark matter masses in the range keV $\lesssim m \lesssim$ GeV have also recently received significant attention. Direct detection of these light dark matter candidates leverages their interactions with various collective excitations or quasi-particles~\cite{Lin:2019uvt,Kahn:2021ttr}. In order to obtain robust bounds, such interactions are best modeled using an EFT framework. Interesting work in this direction has been done for superfluid Helium~\cite{Schutz:2016tid,Knapen:2016cue,Acanfora:2019con,Caputo:2019cyg,Baym:2020uos}, superconductors~\cite{Hochberg:2015pha}, magnetic~\cite{Trickle:2019ovy,Trickle:2020oki,Mitridate:2020kly}, and Dirac~\cite{Coskuner:2019odd} materials, and carbon structures~\cite{Hochberg:2016ntt,Cavoto:2017otc}, to name a few. See also \cite{Mitridate:2022tnv} for more details. Proposals for testing instead the ultralight regime ($10^{-22}$ eV $\lesssim m \lesssim 10^{-15}$ eV) as well as the dark energy sector using cold atoms have also been put forward~\cite{Arvanitaki:2014faa,Hamilton:2015zga}. Taken all together, these efforts underscore the importance of modeling both the dark sector and the excitations of materials used for its detection in a unified, EFT-based language.

\section*{Acknowledgments}

We would like to thank A. Esposito, P. Gorantla, H.T. Lam, and N. Seiberg for comments on the manuscript. The work of TB is supported in part by the grant No.~PR-10614 within the ToppForsk-UiS program of the University of Stavanger and the University Fund. SAH is supported by Simons Investigator award \#620869 and STFC consolidated grant ST/T000694/1. PK is supported in part by the NSERC of Canada. HL is supported by the Office of High Energy Physics of US DOE under grant Contract Number DE-SC0012567 and DE-SC0020360 (MIT contract \#578218). MM is
supported in part by the Simons Foundation grant 488657 (Simons Collaboration on the
Non-Perturbative Bootstrap) and the BSF grant no.~2018204.
The work of RP is supported in part by the National Science Foundation under Grant No. PHY-1915611.  AN is partially supported by the US DOE (award number DE-SC011941) and by the Simons Foundation (award number 658906). DTS is supported in part by the US DOE grant No. DE-FG02-13ER41958, by a Simons Investigator grant, and by the Simons Collaboration on Ultra-Quantum Matter, which is a grant from the Simons Foundation (award number 651440).


\newpage

\bibliography{Biblio}
\bibliographystyle{utphys}

\end{document}